\documentclass[conference]{IEEEtran}
\IEEEoverridecommandlockouts
\usepackage{cite}
\usepackage{amsmath,amssymb,amsfonts}
\usepackage{algorithmic}
\usepackage{graphicx}
\usepackage{textcomp}
\usepackage{xcolor}

\def\BibTeX{{\rm B\kern-.05em{\sc i\kern-.025em b}\kern-.08em
    T\kern-.1667em\lower.7ex\hbox{E}\kern-.125emX}}
\begin{document}

\title{Transfer Learning-Enhanced Instantaneous Multi-Person Indoor Localization by CSI\\
{\footnotesize }
\thanks{This work was supported by the National Natural Science Foundation of China (Grant No. 62071192) and the National Key Research and Development Program of China (Grant No. 2020YFB1806904).}
}
\author{\IEEEauthorblockN{1\textsuperscript{st} Zhiyuan He, 2\textsuperscript{nd} Ke Deng, 3\textsuperscript{rd} Jiangchao Gong, 4\textsuperscript{th} Yi Zhou, 5\textsuperscript{th} Desheng Wang}
\IEEEauthorblockA{\textit{Wuhan National Laboratory for Optoelectronics} \\
\textit{Huazhong University of Science and Technology}\\
Wuhan,  China \\
1\textsuperscript{st} zedyuanhe@hust.edu.cn, 2\textsuperscript{nd} sheepsui@hust.edu.cn, 3\textsuperscript{rd} m202272478@hust.edu.cn\\
4\textsuperscript{th} u202014002@hust.edu.cn, 5\textsuperscript{th} dswang@hust.edu.cn}
}

\maketitle

\begin{abstract}
Passive indoor localization, integral to smart buildings, emergency response, and indoor navigation, has traditionally been limited by a focus on single-target localization and reliance on multi-packet CSI. We introduce a novel Multi-target loss, notably enhancing multi-person localization. Utilizing this loss function, our instantaneous CSI-ResNet achieves an impressive $99.21\%$ accuracy at $0.6m$ precision with single-timestamp CSI. A preprocessing algorithm is implemented to counteract WiFi-induced variability, thereby augmenting robustness. Furthermore, we incorporate Nuclear Norm-Based Transfer Pre-Training, ensuring adaptability in diverse environments, which provides a new paradigm for indoor multi-person localization. Additionally, we have developed an extensive dataset, surpassing existing ones in scope and diversity, to underscore the efficacy of our method and facilitate future fingerprint-based localization research.
\end{abstract}

\begin{IEEEkeywords}
Wireless sensing, Channel state information(CSI), WIFI localization, Fingerprinting, Signal Processing
\end{IEEEkeywords}

\section{Introduction}
Amidst the technological evolution from 5G towards 6G, passive indoor localization technologies independent of specialized portable equipment have attracted substantial attention within the academic and industrial spheres. These technologies underpin a wide array of location-based services and applications, extending to elderly monitoring\cite{b3}, intrusion detection\cite{b2}, and real-time human tracking\cite{b5}. Despite certain advancements made by vision-based methods\cite{b6} in this domain, they confront dual challenges posed by lighting conditions and privacy concerns. Similarly, prevalent infrared positioning suffers from accuracy issues influenced by environmental temperatures\cite{b19}. In contrast, traditional radar-based systems are often prohibitive in cost for widespread deployment\cite{b8}. In this context, WiFi-based positioning systems, leveraging their extensive availability and cost-effectiveness, have become a preferred solution to address these challenges\cite{b10}.

Although some scholars concentrate on utilizing Angle of Arrival (AoA) and Time of Flight (ToF) for localization, the resolution of these techniques is often constrained due to the limited bandwidth and antenna count of commercial WiFi devices\cite{b8,b11}, thereby narrowing their practical applicability. To surmount these challenges, recent scholarly efforts have pivoted towards WiFi fingerprint-based positioning methods\cite{b15}. In earlier phases of research, the emphasis was on utilizing Received Signal Strength (RSS) for fingerprint characterization. However, in Non-Line-of-Sight (NLOS) conditions, RSS demonstrates increased stochasticity due to complex interactions involving reflections, diffractions, and refractions\cite{b13}. In contrast, phase measurements exhibit periodic variations correlated with propagation distances, thus providing a more robust metric. Additionally, the phase information within wireless links encapsulates pertinent data in channel responses, making the combined use of phase and amplitude information a richer and more reliable approach for localization purposes\cite{b16}.

In this research, we focus on developing an instantaneous, passive indoor localization methodology that integrates both amplitude and phase information. Employing 802.11ac devices, which provide rich carrier and antenna dimensionality, we collect Channel State Information (CSI). A meticulous analysis of the amplitude-phase characteristics of CSI is conducted for efficient data preprocessing, followed by constructing an effective network for localization, which is capable of accurately locating multi-people. The key contributions of this study are outlined as follows:

\begin{itemize}
\item We reconceptualize the localization task as a multi-label prediction problem, introducing an advanced CSI preprocessing strategy to mitigate errors from WiFi restarts, thereby elevating precision.
\item we introduce a CSI-ResNet for localization, leveraging single-timestamp CSI inputs for enhanced accuracy and efficiency in real-time deployment. The network preserves CSI integrity, achieving $99.21\%$ accuracy with a $0.6m$ precision. A nuclear norm-based transfer learning approach is implemented, ensuring adaptability across varied scenarios with minimal initial data.
\item A robust dataset was curated, comprising four experimental scenarios and multi-target data from 60 unique locations. This dataset, excelling in scale and label diversity, significantly advances the groundwork for intricate CSI-based location prediction analyses.
\end{itemize}

This paper is structured as follows: Section \ref{sec2} reviews related research and the study's rationale. Section \ref{sec3} describes the localization system and CSI data preprocessing, including the CSI-ResNet and transfer learning approach. Section \ref{sec4} presents experimental methodologies and analyses. Section \ref{sec5} concludes with a summary of findings and contributions.

\section{Related work}\label{sec2}

\subsection{Indoor Localization}

In the realm of WiFi-based indoor localization, two primary technologies prevail: model-based and fingerprint-based localization\cite{b20}. Model-based approaches typically employ geometric measurements to determine the position relative to several known Access Points (APs). This category includes methods like centroid localization and techniques based on Angle of Arrival (AoA) and Time of Arrival (ToA)\cite{b21,b22}. However, these approaches generally necessitate users to carry communication devices capable of connecting with known APs, which may not always be feasible in certain indoor scenarios.

In contrast to model-based approaches, fingerprint-based methods in WiFi indoor localization incorporate advanced neural network algorithms. Early studies extensively used RSS due to its ease of access in hardware. Despite initial advancements like the Radar system using RSS, the fluctuating nature of RSS values, influenced by multipath fading and shadowing, limits the precision of RSS-based localization, impacting the overall effectiveness of these systems\cite{b23,b24,b25}.

Compared to traditional RSS methods, CSI exhibits significant advantages in indoor localization. CSI provides richer, subcarrier-level information crucial for capturing signal propagation characteristics in complex indoor scenarios. Wu et al.\cite{b26} first applied CSI to fine-grained indoor positioning, establishing a model linking CSI with physical distances and achieving precise positioning through trilateration. Furthermore, Zhou et al. expanded this domain with the HRC algorithm, combining RSS and CSI granularity to create a comprehensive fingerprint database to reduce data dimensions for efficient position estimation\cite{b29}.

Zhang et al.\cite{b30} were pioneers in applying Long Short-Term Memory (LSTM) networks to CSI-based indoor localization, leveraging the temporal features of CSI for precise positioning. Following this, Wang et al.\cite{b16} enhanced the efficiency and accuracy of utilizing temporal features by integrating Convolutional Neural Networks (CNNs) with Bidirectional Gated Recurrent Units (BGRUs). Additionally, Tian et al.\cite{b31} employed Bidirectional LSTM (BLSTM) networks for multidimensional recognition of location and activities, thereby increasing the precision of localization.

\subsection{Motivation}
Current indoor localization research, predominantly single-user oriented, neglects multi-user environment complexities, which leads to accuracy issues in conventional systems due to user characteristic similarities. Furthermore, calibrating phase offsets in WiFi RF chains across diverse deployments remains challenging\cite{b32,b33}. Moreover, high-precision positioning algorithms, reliant on temporal CSI signals, face accuracy challenges due to inherent data instability, such as packet loss\cite{b31}. Additionally, transferability issues are the key to fingerprinting methods.

Our work integrates an automated phase compensation method with a dual-channel CSI-ResNet framework, redefining multi-user positioning into a multi-label prediction task and boosting accuracy. We address temporal CSI instability by using single-packet CSI, reducing disturbances\cite{b31}. Additionally, a nuclear norm-based pre-training strategy is implemented, enhancing our model's swift adaptability and consistent performance in new scenarios.

\begin{figure}[htbp]
\centerline{\includegraphics[width=0.5\textwidth]{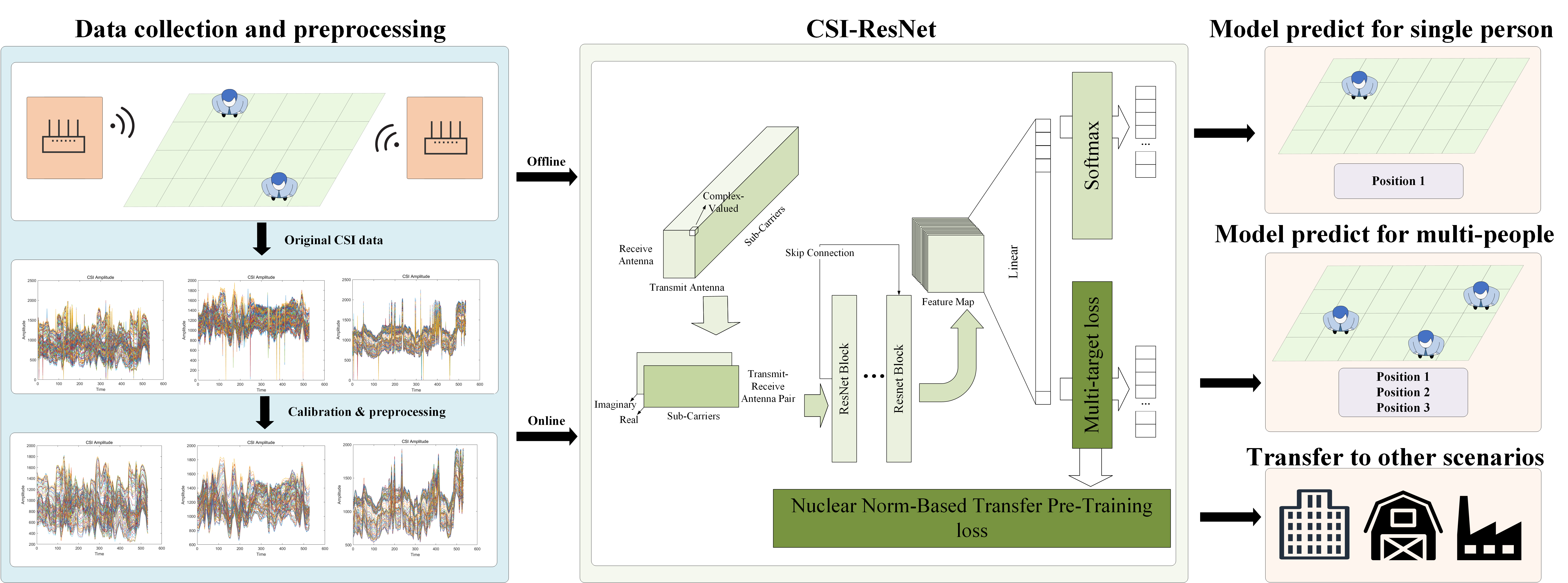}}
\caption{Proposed localization system architecture.}
\label{fig1}
\end{figure}

\section{System model}\label{sec3}
The proposed localization system is illustrated in Fig. \ref{fig1}. During the offline, CSI data are collected and preprocessed to create a fingerprint database, which is then used to train a CSI-ResNet for online multi-person localization.

\begin{figure}[htbp]
\centerline{\includegraphics[width=0.5\textwidth]{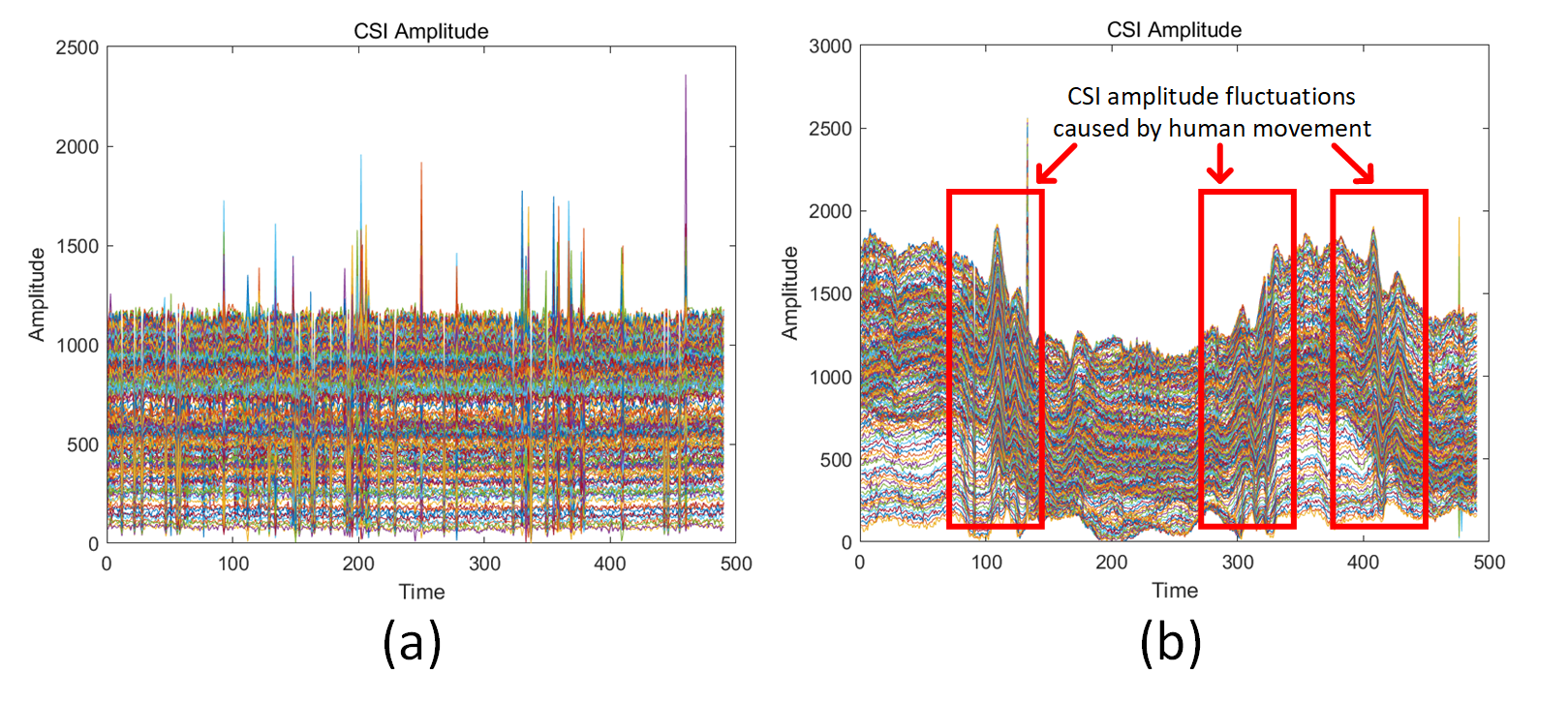}}
\caption{CSI amplitude in two different scenarios. (a) CSI Measured in an Empty Room. (b) CSI Measured in a Room with Occupants.}
\label{fig2}
\end{figure}

CSI signals in wireless channels, as outlined in \cite{b31}, are a blend of static and dynamic multipath elements. Static components stem from structural reflections, while dynamic ones arise from human movements. Our study classifies user characteristics into positional attributes and limb movements. Fig. \ref{fig2} demonstrates that while CSI data reliably indicate position, limb motion reflections are discernible only during activity, emphasizing the distinct effects of human presence and motion on CSI.

Consequently, the received CSI can be mathematically represented as:
\begin{equation}
H(f,t) = H_s(f,t) + H_m(f,t) + H_n(f,t)
\end{equation}

Where $H_s(f,t)$ denotes the CSI measurements transmitted by static paths such as ceiling/wall reflections, termed as static CSI component. $H_m(f,t)$ and $H_n(f,t)$ represent the CSI measurements transmitted via body reflection paths and limb reflection paths, respectively. $H(f,t)$ can also be expressed as:
\begin{equation}
H(f,t) = \|H(f,t)\| \cdot e^{i\theta}
\end{equation}

Where $\|H(f,t)\|$ and $e^{i\theta}$ symbolize the amplitude and phase of the transmitted signal, respectively. Thus, the received CSI can be rewritten as:

\begin{equation}
\begin{split}
\|H(f,t)\| \cdot e^{j\theta} = \|H_s(f,t)\| \cdot e^{j\theta_s} + \|H_m(f,t)\| \cdot e^{j\theta_m} \\
+ \|H_n(f,t)\| \cdot e^{j\theta_n}
\end{split}
\end{equation}
However, the phase information in CSI is often affected by factors like carrier frequency offset and sampling frequency offset, leading to significant errors in CSI phase information. Also, the amplitude information in CSI changes with device restarts. Hence, the actual measured CSI is:
\begin{equation}
\begin{split}
\|H_R(f,t)\| \cdot e^{j\theta_R} = \alpha(\|H_s(f,t)\| \cdot e^{j\theta_s}
\\ + \|H_m(f,t)\| \cdot e^{j\theta_m} + \|H_n(f,t)\| \cdot e^{j\theta_n})e^{j\theta_e}
\end{split}
\end{equation}

Where $\|H_R(f,t)\|$ is the actual received amplitude, $\theta_R$ is the actual received phase, $\alpha$ is the amplitude offset, and $\theta_e$ is the intrinsic phase difference between receiving antennas. Therefore, the received CSI requires calibration.

\subsection{CSI preprocessing}\label{AA}

\begin{figure}[htbp]
\centerline{\includegraphics[width=0.5\textwidth]{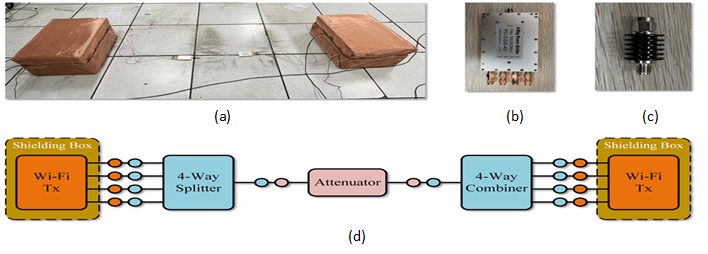}}
\caption{Experiment for CSI calibration. (a) Phase Offset Measurement Experiment Setup. (b) Spliter/combiner. (c) Attenuator. (d) Schematic Diagram of Phase Offset Measurement Experiment.}
\label{fig3}
\end{figure}

 As delineated in research\cite{b32,b33}, frequent rebooting of transceivers impacts oscillator configuration, leading to random phase differences between receiving antennas. For precise intrinsic phase difference acquisition between receiving antennas, we followed the connection scheme depicted in Fig. \ref{fig3} for transceivers and conducted a series of detailed experimental analyses.

\begin{figure}[htbp]
\centerline{\includegraphics[width=0.5\textwidth]{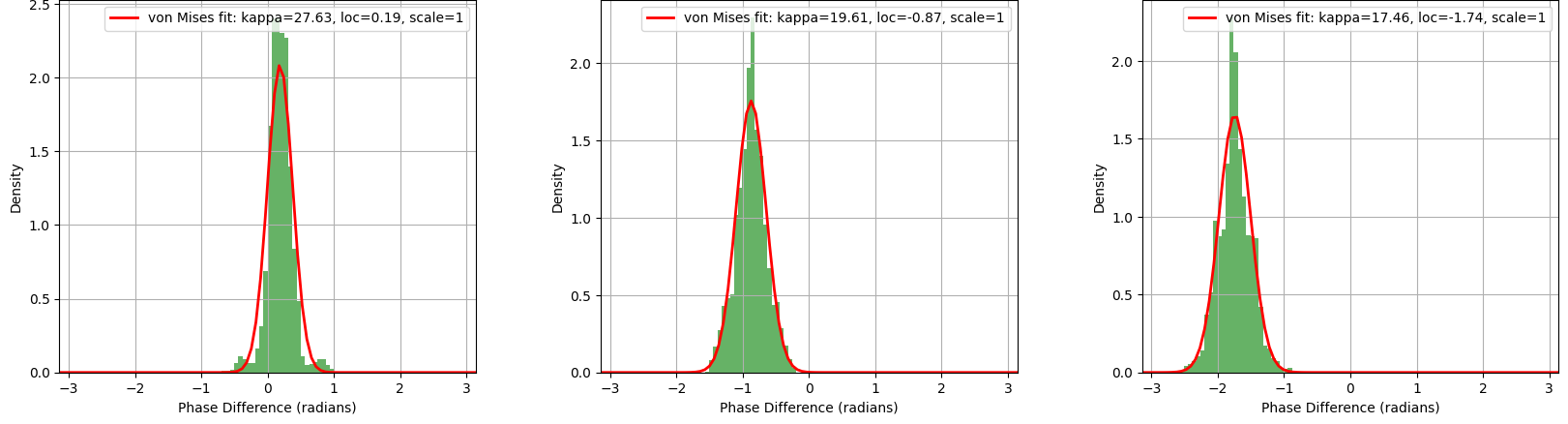}}
\caption{Distribution of phase difference between Rx 2,3,4 with Rx 1.}
\label{fig4}
\end{figure}

After restarting the transceivers 20 times, the phase differences between the receiving antennas were meticulously recorded. By aggregating data across all carriers and time packets, a probability distribution was obtained, as illustrated in Fig. \ref{fig4}. 

Experiments conducted with the RT-AC86U router demonstrate a distinctive pattern in the phase difference between receiving antennas, which is shown in Fig. \ref{fig4}.

Specifically, an integrated analysis of data and fitting the distribution using the von Mises distribution in the angular domain reveals a stable phase difference pattern: between antennas 1 and 2, it's around 0.19 radians (concentration 27.63), between 1 and 3, -0.87 radians (concentration 19.61), and between 1 and 4, -1.74 radians (concentration 17.46). These differences are attributed to inherent antenna properties. Stability is maintained even after system reboots. Consequently, we adjust the CSI for consistent angular offsets and normalize amplitude. The calibrated CSI is represented as:
\begin{equation}
H_{\text{cali}}(M,N) = \frac{| H_{R(M,N)}(f,t) | \cdot e^{j\theta_R} \cdot e^{j\Delta\theta_N}}{\frac{1}{M} \sum_{m=1}^{M} | H_{R(M,N)}(f,t) |}
\end{equation}

Where $M$ and $N$ denote the carrier and receiving antenna, respectively, and $\mathrm{\Delta}\theta_N$ is the compensating phase for the corresponding antenna.

To extract static features related to positioning from CSI data and effectively filter out low-frequency interferences caused by human activities such as walking and breathing, we implemented a preprocessing strategy on CSI data. According to the literature\cite{b35}, the frequency of human walking is approximately 1.9Hz, and the frequency of breathing ranges between 1 to 1.67Hz. To effectively filter out these low-frequency signals, a high-pass filter approach was adopted. Specifically, we applied a Fast Fourier Transform (FFT) to the CSI data in the time domain, followed by the application of a high-pass filter with a 2Hz cutoff. Subsequently, the data was converted back to its original time domain using an Inverse Fast Fourier Transform (IFFT). This process ensures that features extracted from CSI data are focused on reflecting the spatial information and user position characteristics of the positioning scenario, thereby excluding potential interferences from human dynamic activities. The filtered CSI is represented as:
\begin{equation}
\text{HPF}_{>2Hz}(H_{\text{cali}}(f,t)) = H_s(f,t) + H_m(f,t)
\end{equation}

where ${HPF}_{>2Hz}(H_{\text{cali}}(f))$ denotes the high-pass filtered CSI. Additionally, to further enhance the data's accuracy and stability, a Hampel filter was applied to eliminate potential outliers. The preprocessed data serves as the offline input to the network.

\subsection{CSI-ResNet}\label{BB}
CSI analysis reveals that spatial attributes, stable in contrast to the transient nature of human movement data\cite{b31}, align closely with the structured nature of visual data. This similarity underpins the deployment of CSI-ResNet, an advanced adaptation of the ResNet framework, proficient in pinpointing human locational indicators within CSI. ResNet, with its residual learning approach, is particularly effective in spatial feature discernment, overcoming the vanishing gradient challenge in deep neural networks. Meanwhile, GRU and LSTM networks, as discussed in\cite{b16,b31}, are tailored more towards temporal data, making them less ideal for the consistent spatial traits inherent to CSI.

Our enhanced ResNet model employs a bifurcated convolutional input layer, enabling distinct processing of the real and imaginary components of complex CSI data, which adeptly captures amplitude and phase nuances, reconfiguring CSI into a dual-layered framework to better analyze spatial characteristics. By concentrating on the dimensions of transmit-receive antennas, our model gains heightened sensitivity to phase variances, enriching the extraction of locational features. Utilizing CSI's full data range, extending past RSSI, markedly improves the accuracy of indoor positioning. Additionally, the integration of deep residual learning within the ResNet framework refines its capacity for detailed spatial analysis.

\begin{figure*}[htbp]
\centerline{\includegraphics[width=\textwidth]{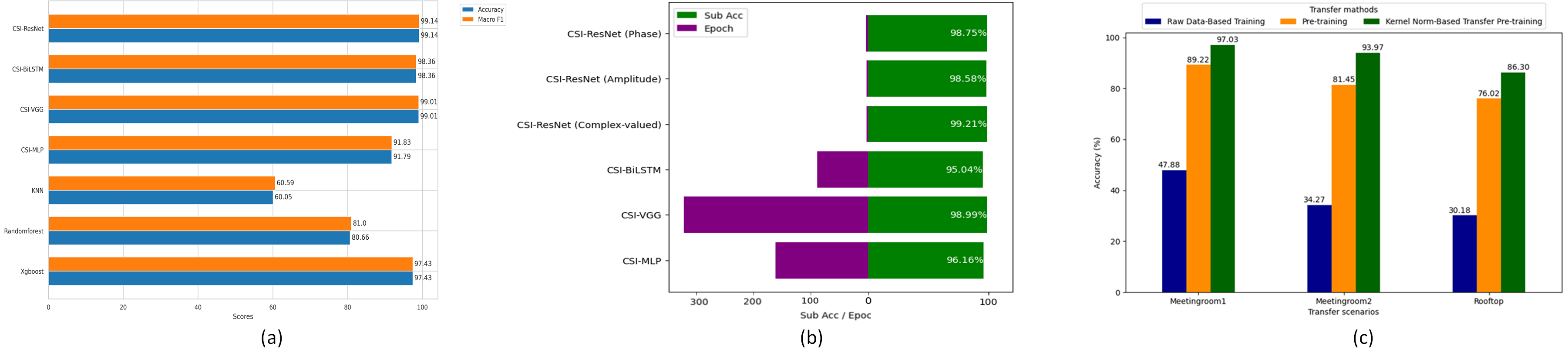}}
\caption{Experimental results. (a) Metrics of different algorithms for single-person positioning. (b) Metrics and epoch of different algorithms for multi-people positioning. (c) Accuracy Comparison Across Different Transfer Scenarios.}
\label{fig7}
\end{figure*}

\subsection{Multi-target loss}\label{CC}
To adapt traditional single-person indoor positioning techniques for multi-person scenarios, a key innovation lies in the optimization of the loss function. Traditional systems typically employ single-label classification methods, identifying one target category from $N$ potential classes. The conventional loss function integrates softmax and cross-entropy, expressed as:
\begin{equation}
L = -\log\frac{e^{s_p}}{\sum_{j=1}^{N}e^{s_j}}
\end{equation}
Here, $s_p$ is the score of the target category, while $s_j$ represents the scores of all candidate categories. This method aims to ensure the highest score for the target category among all classes. However, multi-person indoor localization faces a unique challenge: the discrete nature of individuals in the environment leads neural networks to predict 'empty' scenarios, i.e., all potential target locations as unoccupied. This is primarily due to the prevalence of non-target (unoccupied) samples over target (occupied) samples in training data, causing a bias towards non-target predictions.

In response to the inherent limitations of traditional single-target localization systems in multi-user scenarios, our research innovates by implementing a multi-label classification paradigm. This allows individual samples to be categorized into multiple classes. A novel loss function, termed "Multi-target loss," is designed to ensure that the score of each target category surpasses the score of any non-target category. This methodology effectively recalibrates the predictive balance between target and non-target categories, crucial for accurately capturing sparsely distributed multi-person location data. The loss function is mathematically expressed as:

\begin{equation}
\begin{split}
L_{Multi} = \sum_{i \in P}\max(0, \gamma - s_i) + \sum_{j \in N}\max(0, s_j - \gamma) \\
\approx \sum_{i \in P}\log(1 + e^{\gamma - s_i}) + \sum_{j \in N}\log(1 + e^{s_j - \gamma})
\end{split}
\end{equation}

Where $P$ and $N$ denote the sets of positive and negative classes, respectively. This innovative approach not only enhances the predictive accuracy for target positions but also diminishes the likelihood of falsely predicting unoccupied states, thereby significantly improving the robustness and precision of multi-person localization. The parameter $\gamma$ acts as a critical threshold delineating presence or absence at specific locations, enabling the network to discern between target and non-target positions more sensitively. This advanced loss function enables our model to learn positional information in multi-person scenarios effectively, extending its capabilities beyond predicting a singular most likely position to identifying multiple potential occupant locations, thus substantially enhancing the accuracy and resilience of localization.

Furthermore, to address the challenge of evaluating model performance in multi-target localization scenarios, where predictions may encompass several location labels, we introduce SubACC\cite{b41} as an assessment metric. SubACC is a rigorously exacting measure, typically employed to evaluate a model's proficiency in capturing interdependencies among labels. Its formulation is as follows:

\begin{equation}
\text{SubACC} = \frac{1}{N} \sum_{n=1}^{N} I(\hat{y}_n = y_n)
\end{equation}

Where $y_n$ represents the actual set of position labels for the nth prediction, and $\hat{y}_n$ is the corresponding predicted set. The indicator function $I$ yields a value of 1 if the predicted set $\hat{y}_n$ precisely aligns with the true set $y_n$ and 0 otherwise. This metric stringently evaluates the congruence between predicted and actual positional datasets, providing a comprehensive assessment of the model's performance in multi-label prediction scenarios.

\subsection{Nuclear Norm-Based Transfer Pre-Training}\label{DD}
To enhance robustness and generalizability in diverse indoor localization scenarios through deep CSI data analysis, we implement a novel pre-training paradigm. At the heart of this methodology is the incorporation of nuclear norm regularization into the pre-training loss \cite{b44}, which is:
\begin{equation}
L_{Pre}=L_{Multi}+{\lambda\parallel W\parallel}_\ast
\end{equation}
where $L_{Multi}$ is the multi-target loss mentioned in Section \ref{CC}, ${\parallel W\parallel}_\ast$ represents the nuclear norm of the network weight matrix $W$, and $\lambda$ is the regularization coefficient controlling the influence of the nuclear norm.  In multi-person indoor localization, the model's effectiveness hinges on precise identification and differentiation of individual positions. The nuclear norm, a robust tool for low-rank representation, encourages our network to learn streamlined, low-rank features, thus highlighting key localization attributes. By guiding the network towards low-rank solutions, the model focuses on critical features for accurate multi-person position prediction.

On the other hand, the benefits of low-rank representation in adapting to new scenarios and generalizing are particularly significant in multi-person localization. Simplifying the learned feature space, our model more accurately captures and differentiates human positions, relying on essential signal characteristics. This approach not only accelerates the model's adaptation to new environments but also enhances its versatility in diverse settings, like different office layouts and public spaces. The model's ability to swiftly adapt and optimize using low-rank features streamlines deployment, making it highly practical for dynamic real-world applications.

\section{Experiment}\label{sec4}
\begin{figure}[htbp]
\centerline{\includegraphics[width=0.5\textwidth]{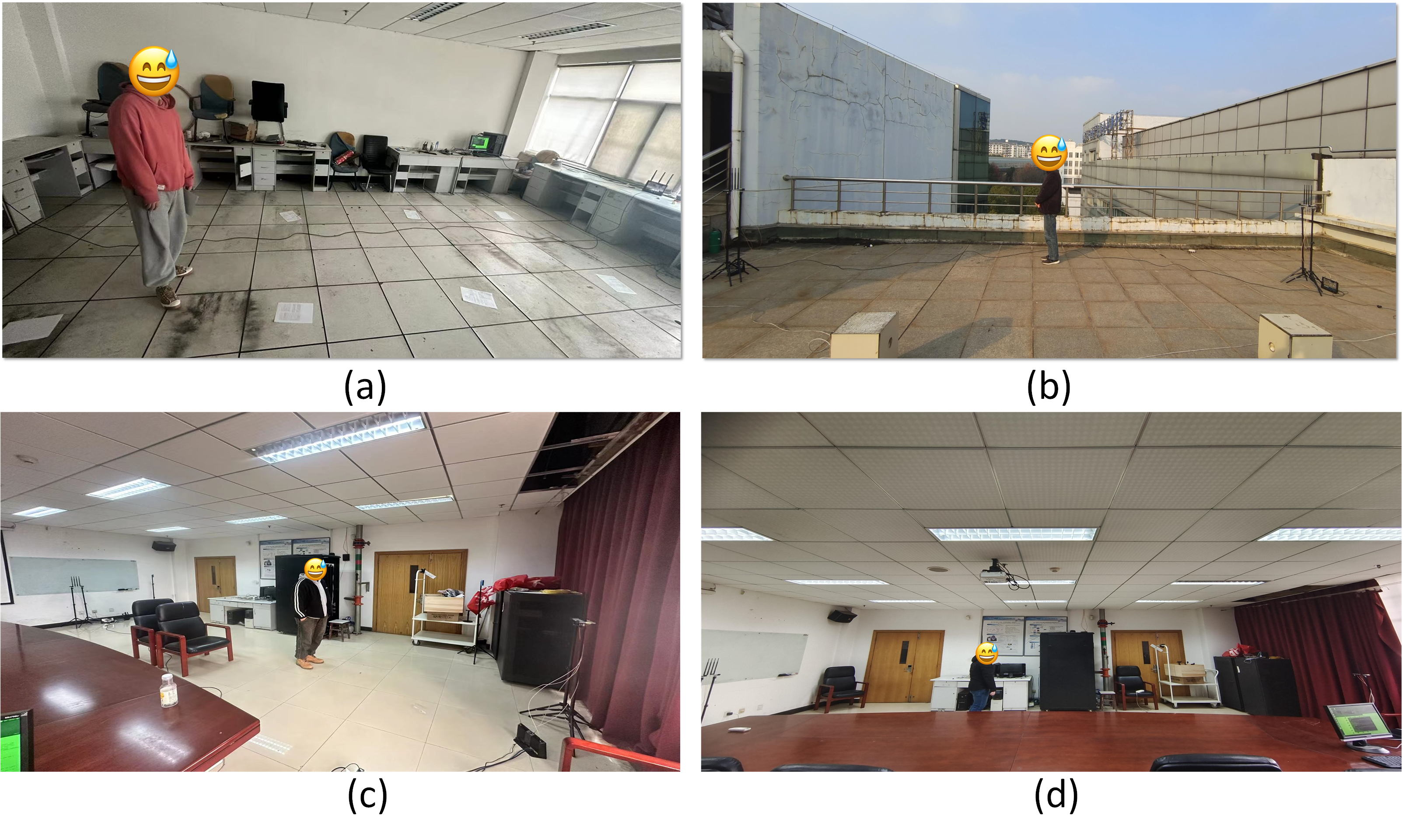}}
\caption{Four experimental scenarios. (a) Laboratory. (b) Rooftop. (c) Meeting room 1. (d) Meeting room 2.}
\label{fig6}
\end{figure}

Our experimental setup encompassed various scenarios: a university lab (Fig. \ref{fig6}(a), $21m^{2}$), an outdoor area for model validation (Fig. \ref{fig6}(b), $40m^{2}$), and two conference rooms (Fig. \ref{fig6}(c)-(d), $35m^{2}$ each). A pair of Asus RT-AC86U routers, modified by Nexmon CSI tool \cite{b42} and controlled via a wired Ubuntu 16.04 host, served as transceivers, situated 1.2 meters high and 8 meters apart. These routers operated on an 80MHz bandwidth at $5.2 GHz$, with each featuring a four-antenna array (9 dB each, 3 cm spacing). Data collection was performed at $50 Hz$, collecting 550 packets per $0.6m$ spaced reference point, aligning with indoor user spatial considerations \cite{b43}.

Proposed CSI-ResNet was evaluated against six state-of-the-art indoor localization algorithms: Random Forest \cite{b38}, MLP \cite{b39}, VGG \cite{b40}, XGBoost \cite{b37}, KNN \cite{b36}, and BiLSTM \cite{b31}. For single-target detection, we used accuracy and Macro F1 score, while SubACC was the metric for multi-person detection. These assessments were based on our extensive dataset, with results presented in Fig. \ref{fig7}.

In single-person localization, CSI-ResNet excelled with a $99.14\%$ accuracy and Macro F1 score, surpassing RF and KNN, as Fig. \ref{fig7}(a) shows. It marginally outperformed other deep learning models like VGG and BiLSTM. XGBoost, despite its potential, fell short in multi-person scenarios due to its lack of multi-label processing. For multi-person localization, CSI-ResNet achieved a $99.21\%$ SubACC, outshining others with just 5 epochs for optimal performance, indicating its rapid deployment efficacy, as depicted in Fig. \ref{fig7}(b). Comparatively, in scenarios utilizing CSI, phase, and RSSI, Fig. \ref{fig7}(b) indicates that while RSSI and phase systems delivered notable results, the CSI-based system reached near $100\%$ accuracy, highlighting CSI's superiority in providing detailed indoor localization data.


Furthermore, CSI-ResNet's transferability was assessed by applying it as a pre-trained model in the scenarios of Fig. \ref{fig6}(b)-(d), initially trained in Fig. \ref{fig6}(a) scenario, using only $1\%$ of the original dataset in new scenarios for initialization. Three transfer training approaches were compared: raw-data-based transfer, pre-trained transfer, and Nuclear Norm-based transfer.

In Fig. \ref{fig7}(c), the results reveal a clear distinction in transfer methods for novel scenarios. Training with raw data fell short, with accuracy under $50\%$ in every scenario, demonstrating incompetence at localization tasks. Pre-trained transfer approach fared better, peaking at $89.22\%$ in a similar setting but dropping to $76.02\%$ in the diverse scenario. However, the Nuclear Norm-based method consistently surpassed $85\%$ accuracy across different scenarios, demonstrating its superior adaptability and transferability in different scenarios. This method's effectiveness in high localization accuracy, only requiring merely $1\%$ of the dataset for initial setup, firmly establishes it as an optimal transfer approach in multi-people localization tasks.

\section{Conclusion}\label{sec5}
Our research heralds a new era in indoor localization with the introduction of a dual-channel CSI-ResNet. This innovative method uniquely converts multi-target challenges into precise multi-label predictions, achieving remarkable precision by utilizing just single-packet CSI data. Incorporating high-pass filtering, phase compensation, and amplitude normalization, we significantly enhance accuracy. The Nuclear Norm-based pre-training strategy ensures outstanding adaptability and generalization, surpassing existing deep learning models with reduced data and training time requirements. Future endeavors will concentrate on advancing user-specific tracking in intricate scenarios, thereby establishing a new standard in indoor localization technology.

\end{document}